\documentclass[aps,prl,twocolumn,prabib,amsmath,amssymb,superscriptaddress,notitlepage]{revtex4}
\usepackage{graphics}
\usepackage{bm}
\usepackage{dcolumn}
\usepackage{natbib}
\usepackage{epstopdf}
\usepackage{float}
\usepackage{graphicx}
\usepackage{epsfig}
\usepackage[pdfstartview=FitH]{hyperref}
\usepackage{color}
\usepackage{appendix}
\usepackage{ulem}

\newcommand{\rr}{{\bf r}}

\begin{document}
\title{Quantum Anomalous Hall Phase Stabilized via Realistic Interactions on a Kagome Lattice}
\author{Yafei Ren}
\affiliation{ICQD, Hefei National Laboratory for Physical Sciences at Microscale, and Synergetic Innovation Center of Quantum Information and Quantum Physics, University of Science and Technology of China, Hefei, Anhui 230026, China}
\affiliation{Department of Physics and Astronomy, California State University, Northridge, California 91330, USA}
\affiliation{CAS Key Laboratory of Strongly-Coupled Quantum Matter Physics, and Department of Physics, University of Science and Technology of China, Hefei, Anhui 230026, China}
\author{T.-S. Zeng}
\affiliation{Department of Physics, The University of Texas at Dallas, Richardson, Texas 75080, USA}
\author{W. Zhu}
\affiliation{Theoretical Division, T-4 and CNLS, Los Alamos National Laboratory, Los Alamos, New Mexico 87545, USA}
\author{D. N. Sheng}
\affiliation{Department of Physics and Astronomy, California State University, Northridge, California 91330, USA}

\date{\today}

\begin{abstract}
We study the quantum phases of spinless fermion at one-third filling on a
Kagome lattice featuring a quadratic band touching Fermi point.
In the presence of weak first and second nearest-neighbor repulsive interactions ($V_1$ and $V_2$), we demonstrate an interaction driven
quantum anomalous Hall effect by employing exact diagonalization and density-matrix renormalization group methods.
The time-reversal symmetry is broken spontaneously by forming loop currents that exhibit long-range correlation.
Quantized Hall conductance corresponding to Chern number of $\pm1$ is obtained by measuring the pumped charge through inserting flux in a cylinder geometry.
We find that the energy gap, which topologically protects the emerging ground states, can be enhanced remarkably by
a moderate $V_2 < V_1$ via calculating the spectrum and charge excitation gaps,
which highlights the experimentally feasible scheme of realizing interaction driven topological phase by
spatially decaying interactions on topologically trivial lattice models.
\end{abstract}

\maketitle

\textit{Introduction.---}
As an analogy of quantum Hall effect discovered in the presence of strong perpendicular magnetic fields, the quantum anomalous Hall effect (QAHE) was first proposed by Haldane for honeycomb lattice with staggered magnetic flux breaking time-reversal symmetry~\cite{QAHE_Haldane_88}. Such QAHEs have been widely explored in non-interacting systems where nontrivial band topology arise from magnetization and spin-orbit coupling~\cite{kane, Liu2016, QAHE_TIthinfilm_10, QAHE_Graphene_Qiao_10}.
Meanwhile, in strongly interacting systems, the searching of exotic ground states has stimulated the interests of interaction-driven QAHE from topologically trivial bands where spontaneous time-reversal symmetry breaking can be realized by interaction driven loop currents~\cite{QAHE_DiracMott_Raghu_08, Rachel2018}.
The first example was proposed in Dirac band with Hubbard interaction based on mean-field analysis and functional renormalization group theory~\cite{QAHE_DiracMott_Raghu_08}.
Further theoretical studies report controversial results depending on boundary conditions from exact diagonalization (ED)~\cite{QAHE_DiracMott_Herbut_OPC_14, QAHE_DiracMott_ED_13, QAHE_DiracMott_ED_14, QAHE_DiracMott_PRB_15_ED}, whereas more reliable density-matrix renormalization group (DMRG) algorithm suggests that the spontaneous QAHE is predominated by other competing phases~\cite{Motruk2015}.

A key ingredient to realize the interaction-driven QAHE is the presence of Fermi touching point, which also emerges in bands with quadratic crossings protected by both time-reversal symmetry and point group symmetry~\cite{Chong2008}.
This stimulates a broad research interest in various lattice models~\cite{QAHE_QBC_Sun_09, QAHE_QBC_Sun_ColdAtom_12, QAHE_KagomeDecGra_MF_10, QAHE_QBC_Kagome_MF_10, QAHE_Kagome_NN_ColdAtom_18, QAHE_Star_ED_MF_18, QAHE_BLGMott_SU4_10, QAHE_BLGMott_FanZhang_11, QAHE_BLGMott_Science_11, QAHE_Lieb_Tsai_15,  QAHE_QBC_root3_16, QAHE_QBC_CkB_WuHQ_16,QAHE_16_PRL_Zhu,MIT_Kagome_Nishimoto_10}, where mean-field calculations suggest that the QAHE emerges as long as weak repulsive interactions are introduced.
Although some ED results indicate the presence of interaction driven QAHE~\cite{QAHE_QBC_CkB_WuHQ_16, QAHE_Star_ED_MF_18}, those evidences, i.e., double degeneracy of ground state and finite loop currents for small systems, are subjected to the finite size effect.
Solid numerical evidences are demanded to confirm whether these phases are stable against quantum fluctuation in the thermodynamic limit.

Recently, several numerical evidences of interaction-driven QAHE are reported on both Kagome~\cite{QAHE_16_PRL_Zhu} and checkerboard~\cite{Zeng2018,Sur2018} lattices based on the state of the art  DMRG studies.
The story for the Kagome systems turns out to be interesting. Up to third nearest-neighbor interactions with comparable strengths are required in the lattice model to realize strong and robust QAHE with a larger excitation gap, which makes it difficult to be realized experimentally.
Moreover, in contrast to the mean-field results where infinitesimal interaction can induce QAHE instability, finite interactions may be required to stabilize the QAHE~\cite{QAHE_16_PRL_Zhu}, leaving the physics in the weak interaction regime unsettled.  
With only first nearest-neighbor hopping and interaction on Kagome lattice, earlier DMRG results suggest that the system remains
a metal for weak interaction~\cite{MIT_Kagome_Nishimoto_10}, which is in contrast to mean-field results~\cite{QAHE_KagomeDecGra_MF_10, QAHE_QBC_Kagome_MF_10}. The main goals of this work are to address the fate of the system in the presence of weak interactions, and to explore more realistic conditions for realizing QAHE for potential experimental systems.

In this work, we numerically map out the quantum phase diagram driven by first and second nearest-neighbor repulsive interactions $(V_1,V_2)$ at one-third filling of spinless fermion on Kagome lattice through ED and DMRG simulations. 
Our extensive ED calculations demonstrate the emergence of QAHE by doubly degenerate ground states and finite loop currents apart from the $V_1$-only case where the energy gap is found to be vanishingly small. 
The presence of $V_2$ interaction enhances the energy gap remarkably, signaling the robustness of the topological phase.
With a finite $V_2 < V_1$, the nontrivial topology is confirmed by large-scale DMRG calculations, which give rise to a uniform circulating loop current spontaneously.
The quantized Hall conductance corresponding to a quantized Chern number $C=\pm1$ is also identified by pumping a unit charge from one side of a cylinder to the other side through inserting $U(1)$ charge flux into the cylinder adiabatically.
When the strength of interactions increase, we reveal a continuous quantum phase transition from the QAHE to a charge density wave (CDW) without any intermediate phase.
Our results provide unbiased numerical evidences of spontaneous QAHE on Kagome lattice stabilized by weak interactions making this model experimentally feasible.

\textit{Model and methods.---}
We consider a spinless fermion-Hubbard model on Kagome lattice with first and second nearest-neighbor interactions. The Hamiltonian is written as
 \begin{align}
H = t\sum_{\langle ij \rangle}c_i^\dagger c_j + V_1\sum_{\langle ij \rangle}n_i n_j + V_2\sum_{\langle \langle ij \rangle \rangle}n_i n_j,
\label{Ham}
\end{align}
where $c_i^\dagger$ ($c_i$) is the creation (annihilation) operator of a fermion at site $i$ and $n_i = c_i^\dagger c_i$ is the particle number operator.
$V_1$ and $V_2$ are the strengths of repulsive interactions between the first ($\langle\ldots\rangle$) and second ($\langle\langle\ldots\rangle\rangle$) nearest neighbors.
We focus on the one-third filling case in a finite system of $N_x \times N_y$ unit cells with total number of sites $N_s = 3\times N_x \times N_y$ and the number of fermions $N_e = N_s/3$.
Here, we take $t = 1$, in which the lower energy band is flat having a quadratical crossing with the middle one.
 
To characterize the topological property of the ground states driven by interactions, we employ the DMRG algorithm combined with ED method.
In ED calculations, we study systems up to 36 sites. 
In a periodic torus geometry, the energy eigenstates can be labeled by the total momentum $\bm{k}=(k_x,k_y)$.
To explore larger systems, we exploit finite or infinite DMRG on either periodic torus or cylinder geometry 
where the boundary is open (periodic) along $x$ ($y$)-direction.
In DMRG calculations, we set $N_y$ up to 5 unit cells (15 lattice sites) and keep the DMRG states up to $M=4800$ to guarantee a good 
convergence (with the truncation error around $10^{-5}$).

\begin{figure}
  \includegraphics[height=2.8in,width=3.4in]{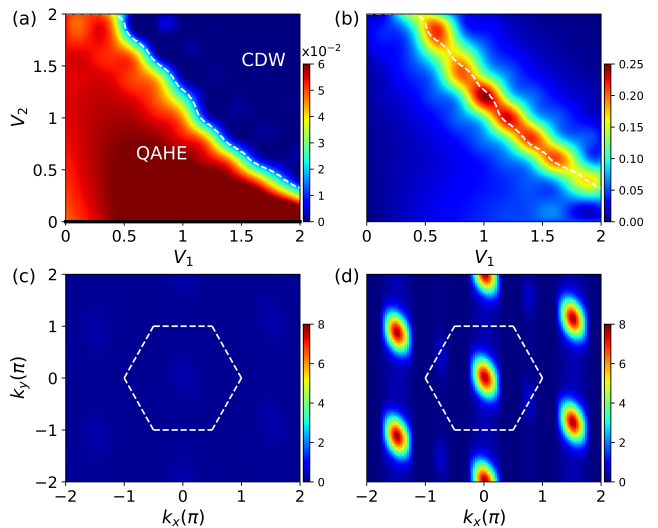}
  \caption{Phase diagram from ED calculation on $N_s = 3 \times 4 \times 3$ tours based on color maps of (a) the loop current amplitude and (b) energy difference between lowest two states. The phase boundary is indicated by a white dashed line. The thick black line on $V_1$-axis indicates that the quantum phase of $V_2=0$ is undetermined due to the vanishingly small energy gap. Contour plots of static density structure factors $S({\bm k})$ are shown for (c) QAHE and (d) $q=(0,0)$ CDW phase. White dashed lines indicate the first Brillouin zone.}
  \label{PhaseDiagram}
\end{figure}

\textit{Phase diagram.---}
In the presence of interactions, we map out the quantum phase diagram in the parameter space spanned by $V_1$-$V_2$, based on both bond current and charge density orders.
Two topologically distinct phases, i.e., QAHE and CDW phases, are displayed in Fig.~\ref{PhaseDiagram}(a).
The QAHE phase is characterized by doubly near degenerate ground states $|\psi_\pm \rangle$. We make complex
superposition of these lowest two states, which possess opposite chiralities and are related to each other by time-reversal operation.
Such a near degeneracy is reflected by a small energy difference between the lowest two energy levels as shown in Fig.~\ref{PhaseDiagram}(b).
For a complex superposition state from these lowest energy states, the time-reversal symmetry is broken by forming loop currents measured by the expectation of current operator
$\langle \hat{j}_{ij}\rangle = i\langle c_i^\dagger c_j - c_j^\dagger c_i \rangle$.
The magnitude of bond current is mapped to color in Fig.~\ref{PhaseDiagram}(a) where a sizable current is found in the whole QAHE region.

We point out that a thick black line is plotted for $V_1$ only case, i.e., $V_2=0$, to distinguish it from the QAHE phase.
In this case, although finite loop currents appear, the energy gap protecting the QAHE is found to be vanishingly small,
which is consistent with a gapless state. 
However, the QAHE emerges with the turn on of a weak $V_2$ interaction.
Furthermore, when $V_1$ or $V_2$ increases to cross the phase boundary, the bond current decreases rapidly suggesting 
the transition to a trivial phase.  
The CDW phase exhibits a charge distribution imbalance among different sublattice sites characterized by the density structure factor $S({\bm k})$. As shown in Figs.~\ref{PhaseDiagram}(c) and \ref{PhaseDiagram}(d) for QAHE, $S({\bm k})$ defined for the density correlations of the same sublattice is featureless in the whole Brillouin zone suggesting a uniform density distribution, whereas for the CDW phase, $S({\bm k})$ shows a strong Bragg peak at the center of the first Brillouin zone. 

\begin{figure}
  \includegraphics[height=1.5in,width=3.4in]{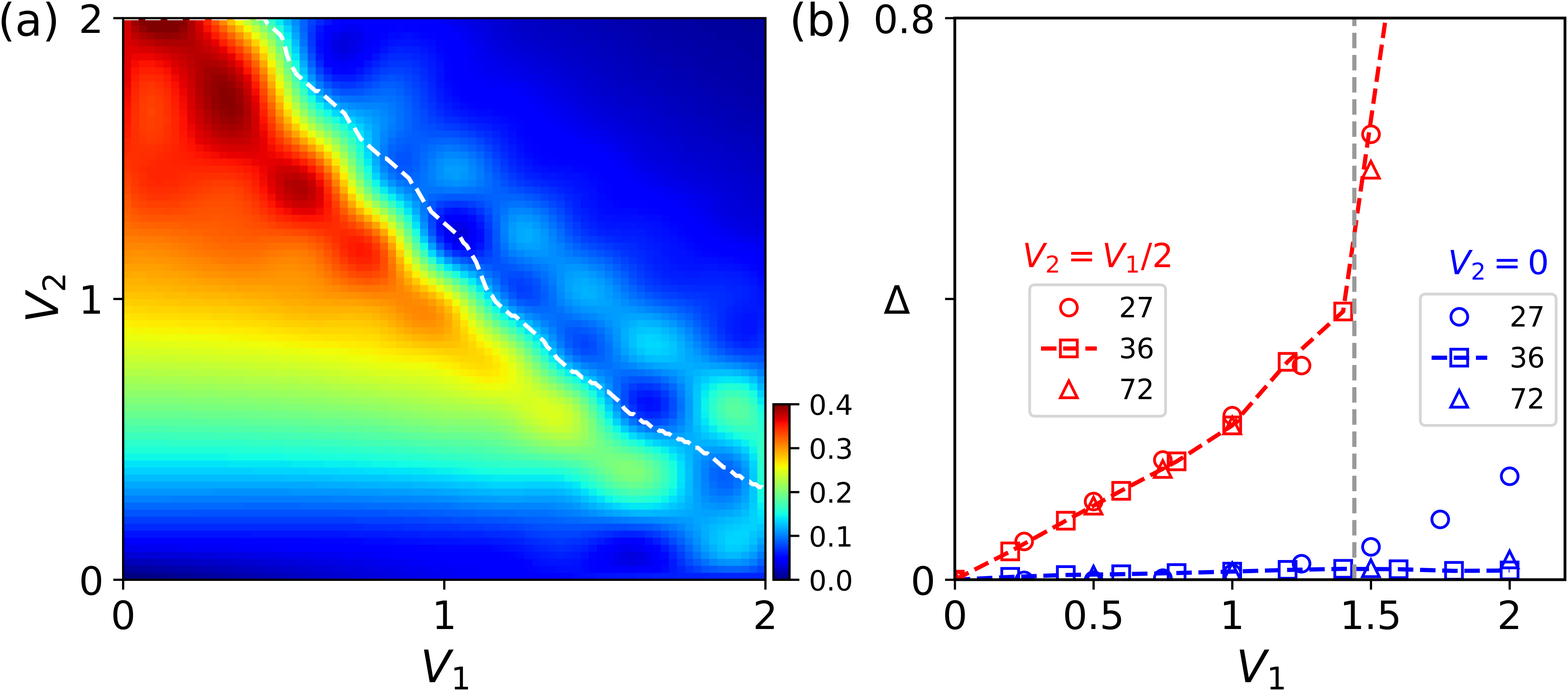}
  \caption{Enhanced energy gap by $V_2$. (a) Color map of spectrum gap between the third and second lowest energy levels calculated by ED method in $N_s=36$ site system. The spectrum gap is vanishingly small for $V_1$ only case and increases significantly in the presence of $V_2$. (b) Charge-hole gap $\Delta(N_s)$ vs $V_1$ for $V_2=0$ and $V_2=V_1/2$ calculated by ED in systems of different sites and by DMRG in system of $N_s=72$ sites. Gray dashed line indicates the critical point between QAHE and CDW phase for $V_2=V_1/2$.}
  \label{ScalingGap}
\end{figure}

\textit{Enhanced gap by second nearest-neighbor interaction.---}
To demonstrate the dependence of energy gap on $V_1$-$V_2$, we show the color map of energy difference between the third and the second lowest energy levels as the spectrum gap in Fig.~\ref{ScalingGap}(a).
Inside the CDW phase, the energy difference is extremely small, manifesting the three-fold degeneracy of the ground state.
In the QAHE phase, however, the spectrum gap exhibits strong dependence on $V_2$.
Extremely small spectrum gap appears near the line of $V_2 = 0$ even for $V_1 = 2$ in agreement with Ref.~\cite{MIT_Kagome_Nishimoto_10}, which reported a gapless metallic phase for $V_1$ only model.
Nevertheless, the presence of $V_2$ can enhance the spectrum gap remarkably.
As shown in Fig.~\ref{ScalingGap}(a), the spectrum gap increases linearly with $V_2$ and shows weak dependence on $V_1$.
A robust gap appears when $V_2$ is moderately large comparable to $t=1$.

Such an enhancement of the gap by $V_2$ is further confirmed by the charge excitation gap $\Delta(N_s)=E_0(N_s, N_e+1)+E_0(N_s, N_e-1)-2E_0(N_s, N_e)$ where $E_0(N_s, N)$ is the ground state energy of the system with $N_s$ sites and $N$ particles.
Figure~\ref{ScalingGap}(b) shows the dependence of $\Delta$ on $V_1$ for different second neighbor interactions $V_2 = 0$ and $V_2 = V_1/2$ for different values of $N_s$.
For $V_2 = 0$, the charge gap $\Delta$ is vanishingly small for different $N_s$ in good agreement with the spectrum gap.
In the presence of finite $V_2 = V_1/2$, $\Delta$ grows up gradually as $V_1$ increases, signaling the emergence of an incompressible gapped QAHE phase.
The linear dependence of energy gap $\Delta$ on $V_2$ is extraordinary comparing to the exponentially small gap for weak interaction predicted by mean-field 
calculations~\cite{QAHE_KagomeDecGra_MF_10, QAHE_QBC_Kagome_MF_10, QAHE_QBC_Sun_09, QAHE_Kagome_NN_ColdAtom_18}.

\begin{figure}
  \includegraphics[height=1.55in,width=3.4in]{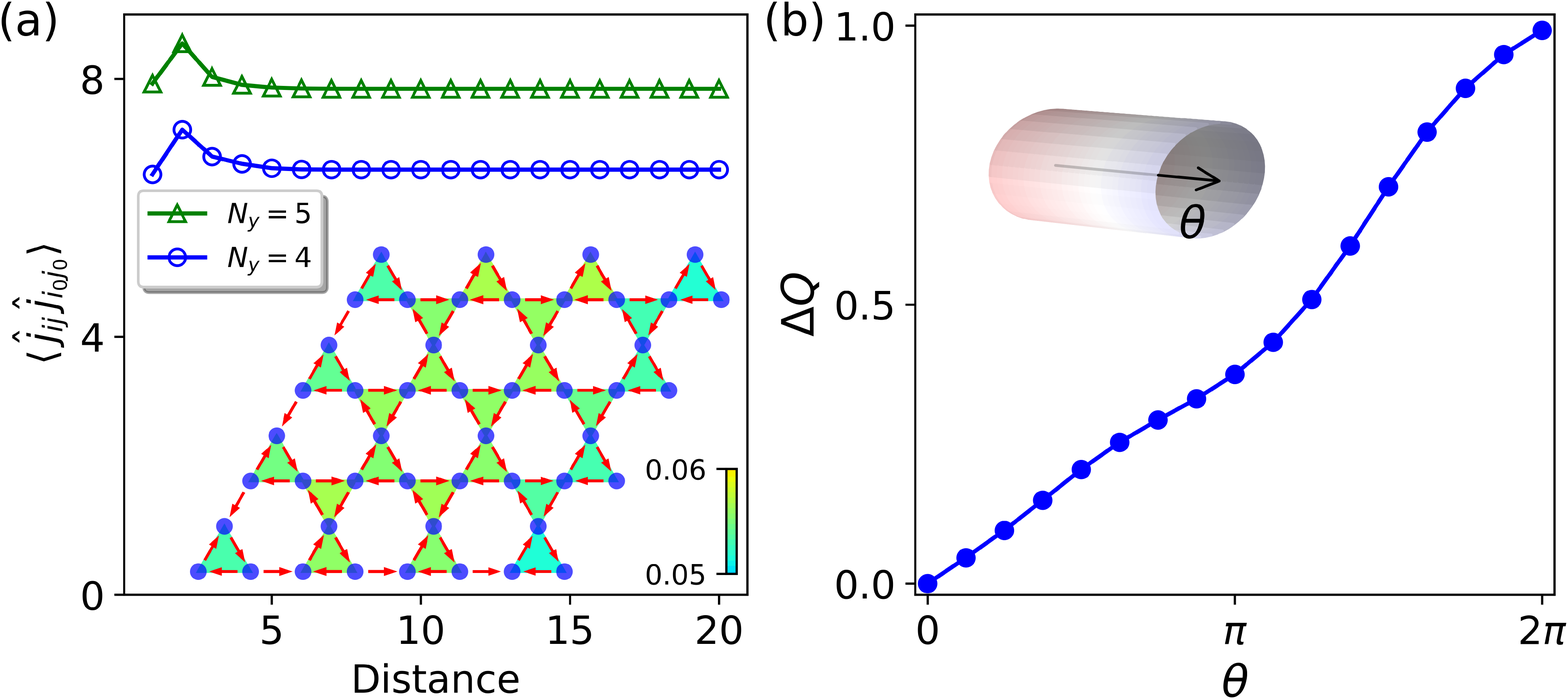}
  \caption{(a) Current-current correlation vs distance for cylinder geometry of $N_y=4$ and $N_y=5$ from infinite DMRG. Finite constant correlation as distance increases indicates the presence of long-range order. Inset shows the current pattern in a $N_s = 3\times 4\times 4$ torus from finite DMRG algorithm. Red arrows indicate the current directions. Color maps the current of each triangle. (b) Net charge transfer $\Delta Q=Q(\theta)-Q(0)$ pumped by threading a $U(1)$ charge flux $\theta$ adiabatically through the hole of a cylinder as illustrated in the inset.}
  \label{CurrentPumping}
\end{figure}

\textit{Current and charge pumping.---}
In relation to the phase diagram, we now perform a numerical DMRG exploration of the QAHE under weak interactions in large systems. In DMRG, we obtain two different ground states $|\psi_{\pm}\rangle$ with near identical energies and opposite loop currents. As these two states are related by time-reversal operation, we focus on $|\psi_+\rangle$ hereinafter. The loop current is schematically shown in the inset of Fig.~\ref{CurrentPumping}(a), which distributes almost uniformly and circulates clockwise (anti-clockwise) around each triangle (hexagon) leading to a vanishing total flux similar to Haldane-honeycomb model~\cite{QAHE_Haldane_88}.
In Fig.~\ref{CurrentPumping}(a), we plot the current-current correlation $\langle \hat{j}_{i,j}\hat{j}_{i_0, j_0}\rangle$ as a function of bond distance $|\rr_{i,j}-\rr_{i_0,j_0}|$ in an infinite cylinder geometry with different widths, where $\hat{j}_{i,j}$ is the current operator between nearest-neighbor sites $i,j$. For different system sizes, the correlation functions tend to converge to finite constants for large distance limit, indicating the behavior of time-reversal symmetry breaking in the thermodynamic limit.

The quantized topological nature of the QAHE is characterized by Chern number $C = 1$ obtained by calculating the topological Laughlin pumping in the $x$-direction by adiabatically inserting a $U(1)$ charge flux $\theta$ into the cylinder hole (as a twist boundary phase in the $y$-direction) based on the recently developed adiabatic DMRG~\cite{Gong2014, Flux_Zhu_15}.
Here, we partition the infinite cylinder along the $x$-direction into two halves. The transverse transfer of total particle number from the right cylinder edge to the left edge is encoded by the variation of the total charge in the left part $Q(\theta)={\rm tr}[\widehat{\rho}_L(\theta)\widehat{N}_{L}]$, where $\widehat{N}_{L}$ and $\widehat{\rho}_L$ are particle number operator and reduced density matrix of left part, respectively. The change $\Delta Q = Q(\theta) - Q(0)$ indicates the transverse transfer of particle as shown in Fig.~\ref{CurrentPumping}(b). In one cycle, a unit of particle $C_{+}=Q(2\pi)-Q(0)\simeq1$ is pumped, visualizing a quantized transverse Hall conductance $\sigma_{xy} = C_{+} e^2/h$ for $|\psi_{+}\rangle$ state.

\begin{figure}
  \includegraphics[height=2.7in, width=3.4in]{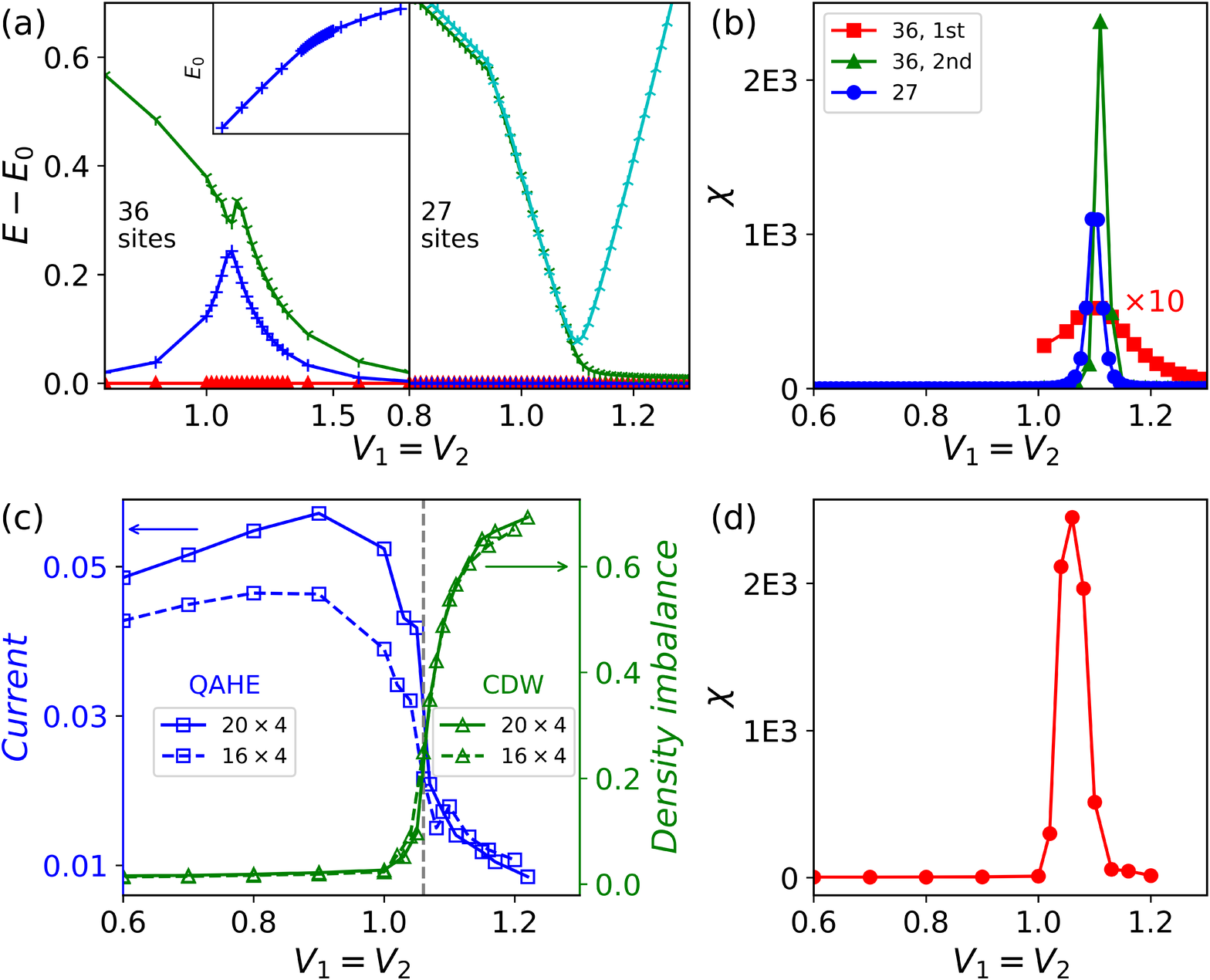}
  \caption{(a) Lowest energy levels vs interaction $V_1=V_2$ for at momentum sector $(\pi, 0)/(0,0)$ for $N_s=36/27$ site system. Inset of $36$-site panel shows the continuous variation of ground state energy as $V_1=V_2$ increases. (b) Fidelity susceptibility $\chi$ vs $V_1=V_2$ for the lowest (red solid square) and second lowest (green solid triangle) energy levels of $N_s=36$ system as well as lowest energy level (blue solid circle) of $N_s=27$ system. (c) DMRG results for bond current magnitude and particle density imbalance between different sublattices in a cylinder of $N_x \times N_y$. Gray dashed line indicates the critical point. (d) DMRG results for fidelity susceptibility of the ground state in a cylinder of $N_x\times N_y=16\times4$.}
  \label{PhaseTransition}
\end{figure}

\textit{Phase transition.---}
In this part, we turn to analyze the phase transition between QAHE and CDW as interaction strength increases.
Without loss of generality, we focus on the case with $V_2=V_1$.
Figure~\ref{PhaseTransition}(a) shows the evolution of the lowest several energy levels at the momentum sector where ground states live.
As the interaction strength increases, the lowest energy level is varying smoothly, and does not show level crossing with excited levels. 
We further calculate the fidelity $f(V)=\langle\psi(V-\delta V)|\psi(V)\rangle$ between two wavefunctions with slightly different interaction strengths.
For two states belong to the same phase, $f(V)$ is close to 1 and the phase transition can be reflected by the peak of fidelity susceptibility $\chi = 2(1-f(V))/\delta V^2$~\cite{Gu2010}. 
As plotted in Fig.~\ref{PhaseTransition}(b), a smooth function $\chi$ with a single peak structure indicates a continuous quantum phase transition from QAHE to CDW, without any evidence of an intermediate phase.

To further verify the continuous transition nature, we also exploit an unbiased DMRG approach for larger systems.
The bond current and maximal density imbalance between different sublattices are presented in Fig.~\ref{PhaseTransition}(c).
Similar to our ED analysis, both order parameters exhibit continuous evolution consistent with a continuous phase transition.
Meanwhile, similar single-peak behavior of $\chi$ is also observed from DMRG calculations in Fig.~\ref{PhaseTransition}(d).
Thus, our ED and DMRG studies consistently support the direct continuous phase transition between QAHE and CDW.

\textit{Experimental realization with cold atom systems.---}
Our results based on $V_1$-$V_2$ model indicate QAHE can be stabilized by the extended repulsive interactions, which is feasible for experimental implementations. With fermionic polar molecules $^{40}\text{K}^{87}\text{Rb}$~\cite{Moses2017} and $^{23}\text{Na}^{40}\text{K}$~\cite{Wu2012}  
loaded into the Kagome optical lattice~\cite{ColdAtomKagome}, the effective interaction potential between fermionic particles
is expected to be in the form $V(\rr-\rr')=d^{2}/|\rr-\rr'|^{3}$ versus distance when the dipole moment $d$ is aligned in the $z$-direction by a strong external field. 
By including up to the third nearest-neighbor interactions and truncating off the tiny terms for longer distance couplings, we confirm that QAHE survives, and verify the robustness of the QAHE from measurements of bond current and energy gap in finite system sizes $N_s=36, 27$ for moderate dipolar interaction strength $d^{2}/a^3\sim t$. Thus our identification of the key role played by weak tails of repulsion interaction suggests that the cold atom trapped polar molecules can naturally realize a QAHE phase on the Kagome lattice.

\textit{Summary and discussion.---}
We have demonstrated a remarkably stable QAHE by neighboring $V_1$-$V_2$ interactions on a Kagome lattice, evidenced by doubly degenerate ground states, spontaneous bond currents with long-range correlation, and quantized Hall conductance.
Without second nearest-neighbor interaction $V_2$, we found vanishingly small spectrum and charge excitation gaps in agreement with previous works.
In the presence of $V_2$, we found the energy gap that protects the ground state exhibits linear dependence on $V_2$, which is strongly enhanced comparing to the exponentially small gap induced by weak interaction $V_1$ predicted by mean-field studies.
By tuning the interactions $V_1$ and $V_2$, the QAHE undergoes a continuous quantum phase transition into a CDW phase. 
Finally, these results imply that the QAHE can be realized by spatially decaying dipolar interactions on a Kagome lattice, making the emergence of the 
topological phase promising within current experimental technologies. Furthermore, our numerical methods 
of identifying such a topological phase can find wide applications for studying interaction driven topological phases 
including quantum spin Hall effect, and QAHE in strongly correlated Mott systems.

\textit{Acknowledgments.---} Y.F.R. acknowledges the financial support from National Key Research and Development Program (Grant Nos. 2016YFA0301700, 2017YFB0405703), the China Government Youth 1000-Plan Talent Program, and the NNSFC(Grant No. 11474265). T.S.Z is supported by Air Force Office of Scientific Research (FA9550-16-1-0387), National Science Foundation (PHY-1505496), and Army Research Office (W911NF-17-1-0128). W.Z. is supported by DOE National Nuclear Security Administration through Los Alamos National Laboratory LDRD Program. D.N.S. is supported by  the DOE, through the Office of Basic Energy Sciences under the grant No. DE-FG02-06ER46305.

\end{document}